\documentclass{kluwer}    
\usepackage{graphicx}

\newdisplay{guess}{Conjecture}

\begin{document}
\begin{article}
\begin{opening}

\title{Eclipsing Binaries With Possible Light-Time Effect \thanks{For further details, please contact
PZ, e-mail address: petr.zasche@email.cz}}

\author{Petr \surname{Zasche}}
\runningauthor{Petr Zasche} \runningtitle{Eclipsing Binaries With
Possible Light-Time Effect} \institute{Astronomical Institute,
Charles University, Prague, Czech Republic}
\author{Miloslav \surname{Zejda}}
\institute{Nicolas Copernicus Observatory and Planetarium, Brno,
Czech Republic}
\author{Lubo\v{s} \surname{Br\'{a}t}}
\institute{Velk\'{a} \'{U}pa 193, Pec pod Sn\v{e}\v{z}kou, Czech
Republic}

\begin{abstract}
The period changes of six eclipsing binaries have been studied
with focus on the Light-time effect. With the least squares method
we also calculated parameters of such effect and properties of the
unresolved body in these systems. With these results we discussed
probability of presence of such bodies in the systems with respect
to possible confirmation by another method. In two systems we also
suggested a hypothesis of fourth body or magnetic activity for
explanation of the "second-order variability" after subtraction of
the light-time effect of the third body.
\end{abstract}
\keywords{stars: binaries: eclipsing -- stars: individual --
multiple stars -- period variations}

\end{opening}

\section{Introduction}
Eclipsing binary systems provide a good opportunity for studying
the presence of an unresolved third body by observing their minima
times because of the light-time effect (hereafter LITE). It was
explained by Irwin \shortcite{Irwin} and its necessary criteria
have been mentioned by Frieboes-Conde \& Hertzeg
\shortcite{Frieboes} and also by Mayer \shortcite{Mayer}. Presence
of the third body in the system is possible only if the times of
minima behave in agreement with a theoretical LITE curve, the
resultant mass function has reasonable value and corresponding
variations in radial velocities are measured. In the last decade
also a confirmation by astrometry seems to be plausible.

In each case we have calculated new light elements of the
eclipsing pair and also the parameters of the predicted third body
orbit. The tables I. and II. present results for each system,
where $M_i$ are masses of components, $p_i$ computed period of the
unresolved body, $A_i$ semiamplitude of LITE, $e_i$ eccentricity,
$\omega_i$ length of periastron, $f(M_i)$ mass function and
$M_{i,min}$ minimum mass (for $i$ = 90$^\circ$) of predicted body,
respectively. The subscript 3 and 4 denotes the parameter of the
third and fourth body, respectively.

\begin{figure}
 \hspace{-2mm}
 \scalebox{0.72}{\includegraphics*[26mm,40mm][196mm,270mm]{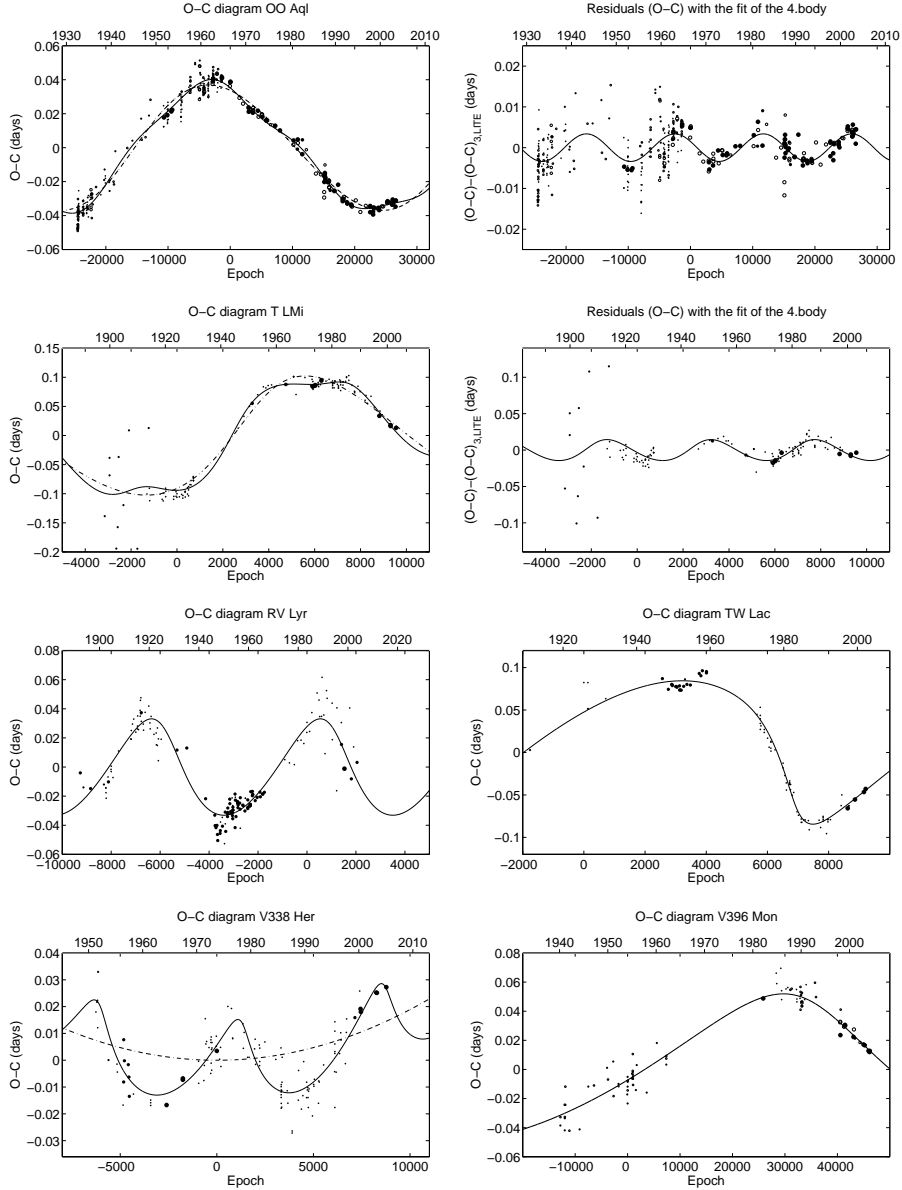}}
 \caption[]{The O-C diagrams of selected eclipsing binaries. The curves
 represent proposed light-time effect caused by the third and fourth
 bodies with periods and semiamplitudes given in the tables I and II. The
 individual primary and secondary minima are denoted by dots and circles,
 respectively. Larger symbols correspond to the photoelectric or CCD
 measurements which were given higher weights in our computations.
 In OO Aql and T LMi cases the LITE caused by the third body is represented
 by the dashdotted line together with the LITE caused by the fourth
 body by the solid line. The residuals after subtraction of LITE caused
 by the third body are shown in figures on the right.}
 \label{fig}
\end{figure}

\section{Investigated systems}
We have derived new linear light elements: \\[1mm]
OO Aql: \( \mathrm{Min\:I} = \mathrm{HJD}\;\:24\;39322.6902 +
0.\!\!^{\rm d}50679191 \cdot {\rm E},\) \\[1mm]
V338 Her: \( \mathrm{Min\:I} = \mathrm{HJD}\;\:24\;41945.3324 +
1.\!\!^{\rm d}30574608 \cdot {\rm E},\) \\[1mm]
T LMi: \( \mathrm{Min\:I}  = \mathrm{HJD}\;\:24\;23856.4200 +
3.\!\!^{\rm d}01988799 \cdot {\rm E},\) \\[1mm]
RV Lyr: \( \mathrm{Min\:I} = \mathrm{HJD}\;\:24\;45526.4003 +
3.\!\!^{\rm d}59901148 \cdot {\rm E},\) \\[1mm]
TW Lac: \( \mathrm{Min\:I} = \mathrm{HJD}\;\:24\;24387.2547 +
3.\!\!^{\rm d}03749292 \cdot {\rm E},\) \\[1mm]
V396 Mon: \( \mathrm{Min\:I} = \mathrm{HJD}\;\:24\;34769.4234 +
0.\!\!^{\rm d}39634326 \cdot {\rm E}.\) \\[-1mm]

In the case of OO Aql and T LMi also the hypothesis of the fourth
body was suggested. The results are in Table II. In the system
OO~Aql the components are very similar and also similar to the Sun
(sp. G5V, T $\simeq$ 5700 K), so the second order variations could
also be caused by the the magnetic cycles, similar to the Sun
(so-called Applegate mechanism, see eg. Applegate
\shortcite{Applegate}).

In the system V338 Her also the quadratic term was applied, so the
mass transfer in the system could be present. From this hypothesis
(with mass transfer parameter $q = 2 \cdot 10^{-10}$) we have
derived the mass transfer rate to $\simeq 4 \cdot 10^{-8}
\mathrm{{M_\odot}/yr}$.

In a few cases here the potential third body would be detectable
in the detailed light curve analysis. Especially the cases where
the third body has not negligible mass, comparing with the
eclipsing pair. Regrettably we have no information about the
distance of individual systems and also about the absolute
magnitudes, so the determination of the angular separation of the
possible third body is very uncertain.

\begin{table}[h!]
 \caption[]{The results for six investigated LITE systems.}
 \label{tabl}
 \scalebox{0.82}{
 \tabcolsep=5pt
 \begin{tabular}{cccccccccc}
 \hline \hline

Name of star & Spectrum  & $ M_1 + M_2$ &  $ p_3 $  &  $A_3$     & $e_3$&$\omega_3$&  $f(M_3) $ & $ M_{3,min} $& Ref. \\
             &           &  [$M_\odot$] &   [years] &  [days]    &      &  [deg]   & [$M_\odot$]&  [$M_\odot$] &  \\
\hline
 OO Aql     &  G5V+G5?   & 1.05+0.88  &    71.72  &   0.0369   & 0.13 &    0.0   &   0.052    &     0.714    & [1] \\
V338 Her    &  F2V+K0    & 1.42+0.40  &    26.35  &   0.0150   & 0.59 &  128.7   &   0.031    &     0.562    & [2] \\
  T LMi     &  A0+G5III  & 6.10+0.60  &   136.98  &   0.1023   & 0.29 &   24.6   &   0.329    &     3.177    & [3] \\
 RV Lyr     &  A5+K4III  & 3.70+1.30  &    67.99  &   0.0331   & 0.31 &  138.5   &   0.044    &     1.192    & [4] \\
 TW Lac     &   A2IV     & 2.80+1.87  &   105.98  &   0.0844   & 0.70 &  213.1   &   0.523    &     3.184    & [5],[2] \\
V396 Mon    &  ?F8+K8?   & 1.20+0.47  &   133.71  &   0.0519   & 0.41 &  114.2   &   0.042    &     0.603    & [6] \\
\hline  \hline
\end{tabular}}
References: [1] -- Al-Naimiy et.al. \shortcite{Naimiy}; [2] --
Budding \shortcite{Budding2}; [3] -- Cester \shortcite{Cester};
[4] -- Budding \shortcite{Budding}; [5] -- Halbedel
\shortcite{Halbedel}; [6] -- Yang \& Liu \shortcite{Yang}.
\end{table}

\begin{table}[t!]
 \caption[]{Parameters for the proposed fourth components.}
 \label{tabl}
 \scalebox{0.82}{
 \tabcolsep=7pt
 \begin{tabular}{ccccccc}
 \hline \hline
Name of star & $ p_4 $  &  $A_4$    & $e_4$ &$\omega_4$&  $f(M_4) $  & $ M_{4,min} $ \\
             & [years]  &  [days]   &       &  [deg]   & [$M_\odot$] &  [$M_\odot$]  \\
\hline
 OO Aql      &   19.71  &   0.0034  & 0.01 &   31.7   &   0.001    &     0.159     \\
  T LMi      &   37.42  &   0.0144  & 0.17 &   44.8   &   0.011    &     1.109     \\
\hline \hline
\end{tabular}}
\end{table}

\section{Conclusion}
We have derived new LITE parameters for six eclipsing binaries by
means of an $O\!-\!C$ diagram analysis. In two cases, OO Aql and
T~LMi, another variation was found, so there is a possibility of a
presence of the fourth body in the system, or magnetic activity in
them. But we have not enough data to make a final decision. So the
consequence is, that for the confirmation of presence of the LITE
in these systems, we need detailed photometric, spectroscopic or
astrometric data of these binaries.

\section{Acknowledgements}
This research has made use of the SIMBAD database, operated at
CDS, Strasbourg, France, and of NASA's Astrophysics Data System
Bibliographic Services. This investigation was supported by the
Czech-Greek project of collaboration RC-3-18 of Ministry of
Education, Youth and Sport and by the Grant Agency of the Czech
Republic, grant No. 205/04/2063

\end{article}
\end{document}